\documentclass[conference]{IEEEtran}
\usepackage[T1]{fontenc}
\usepackage[utf8]{inputenc}
\usepackage{amssymb}
\usepackage[pdftex]{graphicx}
\usepackage{subcaption} 
\usepackage{amsmath}
\usepackage{amsthm}
\usepackage{mathtools}
\usepackage[noadjust]{cite}
\usepackage[hidelinks]{hyperref}
\usepackage[acronym]{glossaries}
\usepackage{siunitx}
\usepackage{tikz}
\usepackage{flushend}
\usepackage{multirow}
\usepackage{pgfplots}
\usepackage{nicefrac}
\usepackage[ruled,boxed,lined,linesnumbered]{algorithm2e}
\pgfplotsset{compat=1.17}

\usetikzlibrary{arrows,chains,matrix,positioning,scopes}
\usetikzlibrary{calc}
\usetikzlibrary{fit,shapes}

\newacronym{NPRACH}{NPRACH}{narrowband physical random-access channel}
\newacronym{ToA}{ToA}{time of arrival}
\newacronym{CFO}{CFO}{carrier frequency offset}
\newacronym{NBIoT}{NB-IoT}{narrowband internet of things}
\newacronym{5GNR}{5G NR}{5G New Radio}
\newacronym{3GPP}{3GPP}{3rd Generation Partnership Project}
\newacronym{UMi}{UMi}{urban microcell}
\newacronym{RMSE}{RMSE}{root-mean-square error}
\newacronym{NN}{NN}{neural network}
\newacronym{BS}{BS}{base station}
\newacronym{UE}{UE}{user equipment}
\newacronym{SG}{SG}{symbol group}
\newacronym{CP}{CP}{cyclic prefix}
\newacronym{OFDM}{OFDM}{orthogonal frequency division multiplexing}
\newacronym{FFT}{FFT}{fast Fourier transform}
\newacronym{AWGN}{AWGN}{additive white Gaussian noise}
\newacronym{DFT}{DFT}{discrete Fourier transform}
\newacronym{FNR}{FNR}{false negative rate}
\newacronym{FPR}{FPR}{false positive rate}
\newacronym{RG}{RG}{resource grid}
\newacronym{RE}{RE}{resource element}
\newacronym{SNR}{SNR}{signal-to-noise ratio}
\newacronym{1D}{1D}{one-dimensional}
\newacronym{MLP}{MLP}{multilayer perceptron}
\newacronym{BCE}{BCE}{binary cross-entropy}
\newacronym{KL}{KL}{Kullback–Leibler}
\newacronym{SGD}{SGD}{stochastic gradient descent}
\newacronym{ppm}{ppm}{parts-per-million}
\newacronym{ICI}{ICI}{inter-carrier interference}
\newacronym{GNN}{GNN}{graph neural network}
\newacronym{BP}{BP}{belief propagation}
\newacronym{FEC}{FEC}{forward error correction}
\newacronym{LDPC}{LDPC}{low-density parity-check}
\newacronym{HDPC}{HDPC}{high-density parity-check}
\newacronym{SCL}{SCL}{successive cancellation list}
\newacronym{SC}{SC}{successive cancellation}
\newacronym{URLLC}{URLLC}{ultra-reliable low-latency communications}
\newacronym{APP}{APP}{a posterior probability}
\newacronym{MIMO}{MIMO}{multiple-input multiple-output}
\newacronym{CNN}{CNN}{convolutional neural network}
\newacronym{BER}{BER}{bit error rate}
\newacronym{BPSK}{BPSK}{binary phase shift keying}
\newacronym{LLR}{LLR}{log-likelihood ratio}
\newacronym{FN}{FN}{factor node}
\newacronym{VN}{VN}{variable node}
\newacronym{CN}{CN}{check node}
\newacronym{MPNN}{MPNN}{message passing neural network}

\newacronym{AI}{AI}{artificial intelligence}
\newacronym{ML}{ML}{machine learning}
\newacronym{SISO}{SISO}{single input single output}
\newacronym{PRB}{PRB}{physical resource block}
\newacronym{PUSCH}{PUSCH}{physical uplink shared channel}

\newacronym{MUMIMO}{MU-MIMO}{multi-user multiple-input multiple-output}
\newacronym{BICM}{BICM}{bit-interleaved coded modulation}
\newacronym{QAM}{QAM}{quadrature amplitude modulation}
\newacronym{LMMSE}{LMMSE}{linear minimum mean square error}
\newacronym{CSI}{CSI}{channel state information}
\newacronym{SIMO}{SIMO}{single-input multiple-output}
\newacronym{CGNN}{CGNN}{convolutional graph neural network}
\newacronym{BLER}{BLER}{block error rate}
\newacronym{LS}{LS}{least squares}
\newacronym{PE}{PE}{positional encoding}
\newacronym{relu}{ReLU}{rectified linear unit}
\newacronym{RB}{RB}{resource block}
\newacronym{CGGNN}{CGGNN}{convolutional graph neural network}
\newacronym{DMRS}{DMRS}{demodulation reference signal}
\newacronym{IoT}{IoT}{internet of things}
\newacronym{ADAM}{ADAM}{adaptive momentum}
\newacronym{TBLER}{TBLER}{transport block error rate}
\newacronym{MCS}{MCS}{modulation and coding scheme}
\newacronym{TDL}{TDL}{tapped delay line}
\newacronym{CDM}{CDM}{code division multiplexing}

\renewcommand{\vec}[1]{\mathbf{#1}}
\newcommand{\vecs}[1]{\boldsymbol{#1}}

\newcommand{\bv}{\vec{b}}

\newcommand{\hv}{\vec{h}}

\newcommand{\mv}{\vec{m}}

\newcommand{\pv}{\vec{p}}

\newcommand{\sv}{\vec{s}}

\newcommand{\wv}{\vec{w}}
\newcommand{\xv}{\vec{x}}
\newcommand{\yv}{\vec{y}}

\newcommand{\zerov}{\vec{0}}

\newcommand{\thetav}{\vecs{\theta}}

\newcommand{\ellv}{\vecs{\ell}}

\newcommand{\Bm}{\vec{B}}

\newcommand{\Hm}{\vec{H}}
\newcommand{\Id}{\vec{I}}

\newcommand{\Mm}{\vec{M}}
\newcommand{\Nm}{\vec{N}}

\newcommand{\Pm}{\vec{P}}

\newcommand{\Sm}{\vec{S}}

\newcommand{\Xm}{\vec{X}}
\newcommand{\Ym}{\vec{Y}}

\newcommand{\Cc}{{\cal C}}

\newcommand{\Nc}{{\cal N}}

\newcommand{\CC}{\mathbb{C}}

\newcommand{\RR}{\mathbb{R}}

\newcommand{\tp}{^{\mathsf{T}}}

\newcommand{\LB}{\left(}
\newcommand{\RB}{\right)}

\newcommand{\LSB}{\left[}
\newcommand{\RSB}{\right]}

\renewcommand{\log}[1]{\mathop{\mathrm{log_2}}\LB #1\RB}

\newcommand\abs[1]{\left| #1 \right|}

\definecolor{mittelblau}{RGB}{0, 126, 198}
\definecolor{violettblau}{cmyk}{0.9, 0.6, 0, 0}
\definecolor{rot}{RGB}{238, 28 35}
\definecolor{apfelgruen}{RGB}{140, 198, 62}
\definecolor{gelb}{RGB}{1, 221, 0}
\definecolor{orange}{RGB}{244, 111, 33}
\definecolor{pink}{RGB}{237, 0, 140}
\definecolor{lila}{RGB}{128, 10, 145}
\definecolor{hellgrau}{RGB}{224, 224, 224}
\definecolor{mittelgrau}{RGB}{128, 128, 128}
\definecolor{dunkelgrau}{RGB}{80,80,80}
\definecolor{anthrazit}{RGB}{19, 31, 31}

\pgfplotscreateplotcyclelist{corporate colours markers}{%
rot, every mark/.append style={fill=.!80!rot},mark=*\\%
mittelblau, every mark/.append style={fill=.!80!mittelblau},mark=square*\\%
apfelgruen, every mark/.append style={fill=.!80!apfelgruen},mark=triangle*\\%
orange, mark=star\\%
pink, every mark/.append style={fill=.!80!pink},mark=diamond*\\%
violettblau, every mark/.append style={fill=.!80!violettblau},mark=otimes*\\%
lila, mark=|\\%
gelb, every mark/.append style={fill=.!80!gelb},mark=pentagon*\\%
hellgrau, mark=text,text mark=p\\%
anthrazit, mark=text,text mark=a\\%
}

\pgfplotscreateplotcyclelist{corporate colours markers double}{%
rot, every mark/.append style={fill=.!80!rot},mark=*\\%
rot, dashed, every mark/.append style={fill=.!80!rot,solid},mark=*\\%
mittelblau, every mark/.append style={fill=.!80!mittelblau},mark=square*\\%
mittelblau, dashed, every mark/.append style={fill=.!80!mittelblau,solid},mark=square*\\%
apfelgruen, every mark/.append style={fill=.!80!apfelgruen},mark=triangle*\\%
apfelgruen, dashed, every mark/.append style={fill=.!80!apfelgruen,solid},mark=triangle*\\%
orange, mark=star\\%
orange, dashed, mark=star\\%
pink, every mark/.append style={fill=.!80!pink},mark=diamond*\\%
pink, dashed, every mark/.append style={fill=.!80!pink,solid},mark=diamond*\\%
violettblau, every mark/.append style={fill=.!80!violettblau},mark=otimes*\\%
violettblau, dashed, every mark/.append style={fill=.!80!violettblau,solid},mark=otimes*\\%
lila, mark=|\\%
lila, dashed, mark=|\\%
gelb, every mark/.append style={fill=.!80!gelb},mark=pentagon*\\%
gelb, dashed, every mark/.append style={fill=.!80!gelb,solid},mark=pentagon*\\%
}

\IEEEoverridecommandlockouts 

\begin{document}

\title{A Neural Receiver for 5G NR Multi-user MIMO}

\author{

\IEEEauthorblockN{Sebastian Cammerer$^1$, Fay\c{c}al A\"{i}t Aoudia$^1$, Jakob Hoydis$^1$, Andreas Oeldemann$^2$, Andreas Roessler$^2$,\\ Timo Mayer$^2$, and Alexander Keller$^1$}
		\IEEEauthorblockA{$^1$NVIDIA, $^2$Rohde \& Schwarz, contact: scammerer@nvidia.com}

\thanks{This work has received financial support from the European Union under Grant Agreement 101096379 (CENTRIC). Views and opinions expressed are however those of the author(s) only and do not necessarily reflect those of the European Union or the European Commission (granting authority). Neither the European Union nor the granting authority can be held responsible for~them.}
\vspace*{-1.0cm}
}

\maketitle

\begin{abstract}
We introduce a \gls{NN}-based \gls{MUMIMO} receiver with \gls{5GNR} \gls{PUSCH} compatibility. The \gls{NN} architecture is based on convolution layers to exploit the time and frequency correlation of the channel and a \gls{GNN} to handle multiple users. The proposed architecture adapts to an arbitrary number of sub-carriers and supports a varying number of \gls{MIMO} layers and users without the need for any retraining.
The receiver operates on an entire 5G NR \emph{slot}, i.e., processes the entire received \gls{OFDM} time-frequency resource grid by jointly performing channel estimation, equalization, and demapping.
The proposed architecture operates less than \SI{1}{dB} away from a baseline using \gls{LMMSE} channel estimation with K-best detection but benefits from a significantly lower computational complexity.
We show the importance of a carefully designed training process such that the trained receiver is universal for a wide range of different unseen channel conditions. Finally, we demonstrate the results of a hardware-in-the-loop verification based on 3GPP compliant conformance test scenarios. 
\end{abstract}
\glsresetall

\section{Introduction}

\Gls{AI} and \gls{ML} have become ubiquitous tools in the wireless research community \cite{honkala2021deeprx,o2017introduction} as well as the telecommunications industry \cite{lin2023artificial}. This trend has been further accelerated by the recent announcement of the \gls{3GPP} consortium to investigate \gls{AI}/\gls{ML} in the physical layer as new study item for the upcoming \gls{3GPP} release 18 \cite{lin2023artificial}. While this will most likely require disruptive changes in the signal structures and the transmitter implementation, our approach takes a different direction and focuses on \gls{NN}-based receivers which are fully compliant with \gls{5GNR}.
We emphasize that neural receivers can be seen as enabling technology for various novel applications such as site-specific basestations that can be re-trained after deployment as well as pilotless communications \cite{aoudia2021end}.

The concept of such a neural receiver was first proposed in \cite{honkala2021deeprx} where the authors have demonstrated a performance close to that of a \emph{classical} receiver with perfect \gls{CSI} in a \gls{SIMO} system by enabling joint channel estimation and data reconstruction. Note that earlier, the authors of \cite{ye2017power} proposed \gls{CNN}-based receiver algorithms for OFDM detection. The neural receiver idea was extended to \gls{MIMO} systems in \cite{korpi2021deeprx}. It was shown later in \cite{aoudia2021end}, that a similar architecture also permits pilotless communications when combined with end-to-end learning \cite{o2017introduction}. Further, neural receivers for \gls{IoT} applications have been demonstrated \cite{soltani2022neural}. However, when considering deep learning for \gls{MUMIMO} systems~\cite{9298921,9103314,scotti2020graph,9771869}, often perfect \gls{CSI} has been assumed, neglecting the issue of implicit channel estimation.
In addition, it is unclear whether evaluations performed on simplistic channel models generalize to more challenging channel environments.

In this work, we extend the concept of neural receivers by a flexible \gls{MUMIMO}
component with \gls{5GNR} \gls{PUSCH} compatibility. A specific focus of the \gls{NN} architecture is on the flexibility with respect to a varying number of users and a configurable number of sub-carriers.
The \gls{NN} architecture itself is a combination of \gls{GNN} and \gls{CNN} layers. Intuitively, the \gls{GNN} offers flexibility regarding varying number of users as it allows an easy reconfiguration similar to \cite{scotti2020graph}.
As a result, approximately 730k trainable weights are required to process an entire \gls{5GNR} slot, meaning the entire time/frequency resources consisting of thousands of resource elements in the time-frequency grid of an \gls{OFDM} system.
The number of \glspl{PRB} is also variable - in fact, our receiver is trained for 4 \glspl{PRB} but evaluated for 217 \glspl{PRB}.
This flexibility is a key enabler for practical deployment as the proposed architecture neither requires any retraining if a different number of users is scheduled, nor if the number of allocated sub-carriers changes.

A practical challenge is to train the system such that it does not overfit to the underlying dataset or channel model.
We therefore randomize the channel during training by using system level \gls{3GPP} \gls{UMi} channel models with random user drops to ensure that the learned receiver generalizes to unseen channel conditions.
The neural receiver and our experiments are implemented using Sionna \cite{hoydis2022sionna}.

\begin{figure*}
    \center
    \resizebox{!}{2.6cm}{\begin{tikzpicture}
\tikzstyle{box} = [draw,rounded corners=.1cm,inner sep=5pt,minimum height=3.1em, text width=5em, align=center,thick]
\def\antenna{%
    -- +(0mm,3.0mm) -- +(2.625mm,6.5mm) -- +(-2.625mm,6.5mm) -- +(0mm,3.0mm)
}
\node[box,draw=apfelgruen] (tx1) at (-6.5,0.8) {Transmitter \\Layer 1};
\node[below=0.15cm of tx1] (txi) {\huge $\dots$};
\node[box,draw=apfelgruen, below=0.6cm of tx1] (tx2) {Transmitter \\Layer $N_T$};

\node [cloud, draw, cloud puffs=12,cloud puff arc=90, thick, aspect=3.5, inner ysep=-0.2em, text width=15em, align=center] (ch) at (0,0) {Channel (3GPP 38.901,\dots)\\ $\yv_{n_F, n_S} = \Hm_{n_F, n_S} \xv_{n_F, n_S} + \wv_{n_F, n_S}$};

\node[box,draw=mittelblau, text width=11em, minimum height=8em] (rx) at (8.0,0) {};

\draw [->,thick]  ([xshift=0.7cm,yshift=0.4cm]tx1.east) -- ([yshift=0.35cm,xshift=-0.04cm]ch.west) node [midway, above=0.15cm, xshift=0.1cm] {$\xv_1$};
\draw [->,thick]  ([xshift=0.7cm]tx2.east) -- ([yshift=-0.35cm,xshift=-0.04cm]ch.west) node [midway, below=0.2cm, xshift=0.3cm] {$\xv_{N_\text{TX}}$};

\draw [->,thick]  ([yshift=0.6cm]ch.east) -- ([xshift=-0.8cm,yshift=1.0cm]rx.west) node [midway, above=0.15cm] {$\yv_1$};
\draw [->,thick]  ([yshift=-0.4cm]ch.east) -- ([xshift=-0.8cm,yshift=-1.0cm]rx.west) node [midway, below=0.3cm] {$\yv_{N_\text{RX}}$};

\node[left=0.6cm of tx1] (b1) {$\bv_1$};
\node[left=0.6cm of tx2] (b2) {$\bv_{N_\text{T}}$};
\draw [->,thick]  (b1) --(tx1) {};
\draw [->,thick]  (b2) --(tx2) {};

\draw[color=apfelgruen,thick] ([yshift=0.4cm]tx1.east) -- ([xshift=0.3cm,yshift=0.4cm]tx1.east) \antenna;
\draw[color=apfelgruen,thick] ([yshift=-0.4cm]tx1.east) -- ([xshift=0.3cm,yshift=-0.4cm]tx1.east) \antenna;
\draw[color=apfelgruen,thick] ([yshift=0.4cm]tx2.east) -- ([xshift=0.3cm,yshift=0.4cm]tx2.east) \antenna;
\draw[color=apfelgruen,thick] ([yshift=-0.4cm]tx2.east) -- ([xshift=0.3cm,yshift=-0.4cm]tx2.east) \antenna;

\draw[color=mittelblau,thick] ([yshift=1.15cm]rx.west) -- ([xshift=-0.4cm,yshift=1.15cm]rx.west) \antenna;
\draw[color=mittelblau,thick] ([yshift=0.4cm]rx.west) -- ([xshift=-0.4cm,yshift=0.4cm]rx.west) \antenna;
\draw[color=mittelblau,thick] ([yshift=-0.4cm]rx.west) -- ([xshift=-0.4cm,yshift=-0.4cm]rx.west) \antenna;
\draw[color=mittelblau,thick] ([yshift=-1.15cm]rx.west) -- ([xshift=-0.4cm,yshift=-1.15cm]rx.west) \antenna;

\node[mittelblau] (sync) at (7.2,1.1) {Neural Receiver};
\node[box,draw=mittelblau, text width=4em, minimum height=1.5em, rotate=90] (sync) at (6.5,0) {Sync.};
\node[box,draw=mittelblau, text width=4em, minimum height=1.5em, rotate=90] (fft) at (7.4,0)  {FFT};
\node[box,draw=mittelblau, text width=4em, minimum height=1.5em, rotate=90] (nn) at (8.3,0)  {NN-RX};
\node[box,draw=mittelblau, text width=1.6em, minimum height=1.5em, rotate=90] (tb1) at (9.3,0.75)  {TB Dec.};

\node[box,draw=mittelblau, text width=1.6em, minimum height=1.5em, rotate=90] (tb2) at (9.3,-0.75)  {TB Dec.};

\node[above=0.75cm of rx.east] (llr11) {};
\node[right=0.6cm of llr11] (llr1) {$\ellv_1$};
\node[below=-0.1cm of rx.east] (llrn1) {};
\node[right=0.1cm of llrn1] (llrn) {\huge $\dots$};
\node[below=0.75cm of rx.east] (llr21) {};
\node[right=0.6cm of llr21] (llr2) {$\ellv_{N_\text{T}}$};

\draw [->,thick] ([xshift=-1.3cm]llr11) -- (llr1) {};
\draw [->,thick] ([xshift=-1.3cm]llr21) -- (llr2) {};

\draw [->,thick, mittelblau] (sync.south) -- (fft.north) {};
\draw [->,thick, mittelblau] (fft.south) -- (nn.north) {};

\draw [->,thick, mittelblau] ([yshift=0.75cm]nn.south) -- (tb1.north) {};
\draw [->,thick, mittelblau] ([yshift=-0.75cm]nn.south) -- (tb2.north) {};

\end{tikzpicture}}
    \caption{Overview of an \gls{MUMIMO} \gls{OFDM} communication system where  $N_\textrm{T}$ \gls{MIMO} layers are received by $N_\textrm{RX}$ antennas.
    \label{fig:sys_overview}}
    \vspace*{-0.4cm}
\end{figure*}
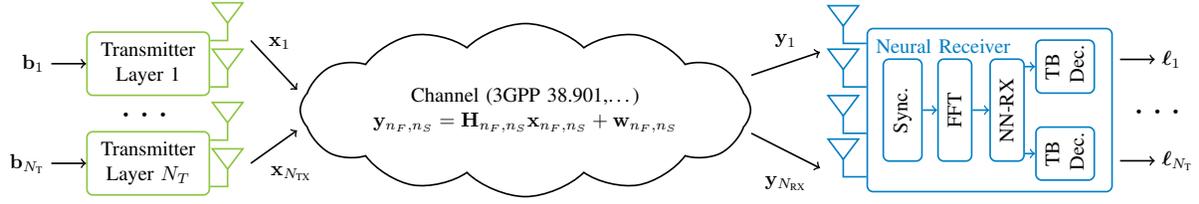

\section{Neural Multi-user MIMO OFDM Receiver}

We consider a \gls{MUMIMO} \gls{OFDM} communication system in which $N_\textrm{T}$ \gls{MIMO} \emph{layers}\footnote{In \gls{5GNR}, each user -- possibly equipped with multiple antennas -- can be configured to transmit multiple \gls{MIMO} \emph{layers} (or data \emph{streams}). For readability, we denote every layer as an independent transmitter in the following description. The proposed neural receiver architecture supports multiple layers per user by treating every stream as a (virtual)~user.} are simultaneously transmitted on the same physical resources to a single receiver with $N_\textrm{RX}$ receive antennas. We assume a total number of $N_\textrm{TX}$ transmit antennas. Such a communication system is illustrated in Fig.~\ref{fig:sys_overview}.

The problem of \gls{MUMIMO} detection is to reconstruct the bits transmitted by the individual layers from the received signal $\yv$.
We denote by $N_\text{S}$ and by $N_\text{F}$ the number of \gls{OFDM} symbols and subcarriers forming the \gls{RG}, respectively.
The $n_\text{T}^{\textit{th}}$ layer ($1 \leq n_\text{T} \leq N_\text{T}$) transmits a vector $\bv_{n_F, n_S, n_\text{T}} \in \{0, 1\}^{m}$ of $m$ bits on every \gls{RE} $\left[ n_F, n_S \right]$ allocated to data transmission, where $1 \leq n_F \leq N_\text{F}$ and $1 \leq n_S \leq N_\text{S}$.
To that aim, every vector $\bv_{n_F, n_S, n_\text{T}}$ is mapped onto a complex-valued baseband symbol denoted by $\xv_{n_F, n_S, n_\text{T}} \in \CC$ which is typically done by using a $2^m$ \gls{QAM} with Gray labeling. \Gls{MIMO} precoding can be applied if a user has more than one transmit antenna.
This leads to the \gls{OFDM} \gls{RG} of baseband modulated symbols
\vspace*{-0.4cm}
\begin{equation}
    \Xm_{n_\text{T}} =
    \begin{bmatrix}
        \xv_{1,1,n_\text{T}} & \cdots & \xv_{1,N_\text{S},n_\text{T}} \\
        \vdots & \ddots & \vdots \\
        \xv_{N_\text{F},1,n_\text{T}} & \cdots & \xv_{N_\text{F},N_\text{S},n_\text{T}}
    \end{bmatrix}.
\end{equation}
Some positions in $ \Xm_{n_\text{T}}$ are not used for data transmission but send known pilots, so-called \gls{DMRS} pilots.
We denote the set of pilot positions as $\mathcal{P}_{n_\text{T}}$ for each layer $n_\text{T}$ where each $\pv_{n_\text{T}} \in \mathcal{P}_{n_\text{T}}$ is given as
$\pv_{n_\text{T}}=\left[{p}_{\text{F}}, {p}_{\text{S}} \right] \in \mathbb{N}^{2}$ defining the indices of pilot positions in the~\gls{RG}.

We assume \gls{OFDM} with a sufficiently long \gls{CP} duration.
Hence, the received \gls{RG} is denoted by $\Ym = \{ \yv_{n_F, n_S} \}_{{1 \leq n_F \leq N_\text{F}, 1 \leq n_S \leq N_\text{S}}}$, where $\yv_{n_F, n_S} \in \CC^{N_\text{RX}}$ is the received signal for the \gls{RE} $\left[ n_F, n_S \right]$ and given as
\begin{equation}
    \label{eq:ch}
    \yv_{n_F, n_S} = \Hm_{n_F, n_S} \xv_{n_F, n_S} + \wv_{n_F, n_S}
\end{equation}
where $\xv_{n_F, n_S} = \LSB x_{n_F, n_S, 1}, \dots, x_{n_F, n_S, N_\text{TX}} \RSB\tp \in \CC^{N_\text{TX}}$ is the vector of transmitted baseband symbols, $\Hm_{n_F, n_S} \in \CC^{N_\text{RX} \times N_\text{TX}}$ is the channel matrix, and $\wv_{n_F, n_S} \sim \Cc\Nc(\zerov, \sigma^2 \Id_{N_\text{RX}})$ is the complex-valued \gls{AWGN} with noise power $\sigma^2$.
Additional effects such as hardware impairments could be easily integrated in this model.

\subsection{5G NR PUSCH terminology and structure}
\label{sec:5gnr}
\begin{figure}
    \centering
    \resizebox{0.64\columnwidth}{!}{\includegraphics{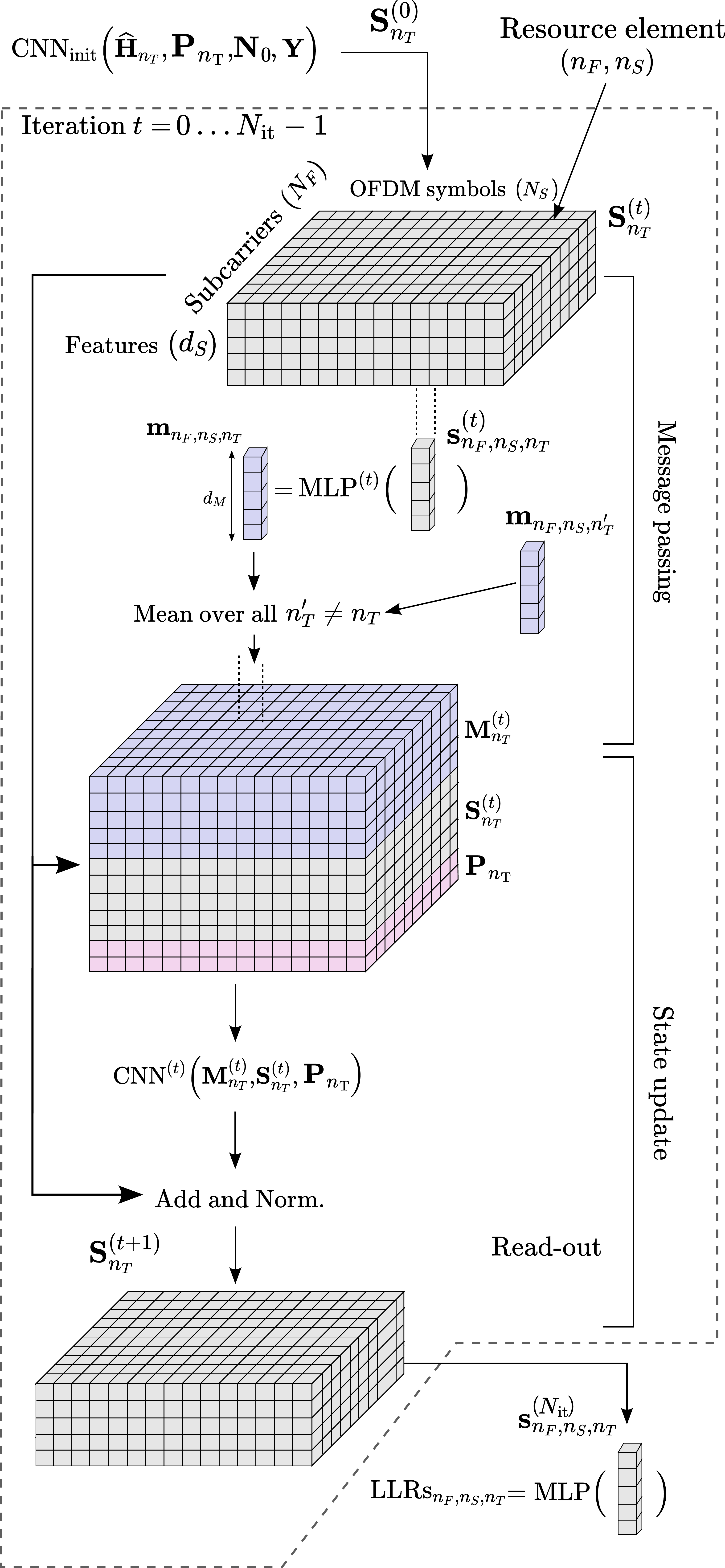}}
    \caption{Architecture of the neural receiver used for MU-MIMO detection. The detection process is shown for layer $n_\text{T}$. Information from other layers $n'_T$ is considered
    in the message passing step.\label{fig:nn}}
\end{figure}

We introduce the \gls{5GNR} related terminology required to understand the following receiver implementation. For further details, we refer the interested reader to \cite{dahlman20205g}.\footnote{An interactive \gls{5GNR} \gls{PUSCH} tutorial is available online \emph{https://nvlabs.github.io/sionna/examples/5G\_NR\_PUSCH.html}.}
In \gls{5GNR}, each  \emph{frame} comprises 10 subframes, each of duration \SI{1}{\milli\s}. 
Each one of these subframes is partitioned into \emph{slots}, each consisting of 12 or 14 \gls{OFDM} symbols. The number of slots depends on the \emph{numerology} defining the subcarrier spacing which takes values ranging from \SI{15}{\kilo\Hz} to \SI{120}{\kilo\Hz}, e.g., for \SI{30}{\kilo\Hz} each frame consists of 20 slots.
The subcarriers are grouped into \glspl{PRB} where each \gls{PRB} denotes 12 subsequent subcarriers.
Further, the so-called \gls{DMRS} denotes the pilot symbols for channel estimation at pre-defined positions in each slot (see \cite[Chapter~9.11]{dahlman20205g}). We configure the system such that the \gls{DMRS} occurs at \gls{OFDM} symbol 2 and 11. 
Each layer is assigned an individual \gls{DMRS} \emph{port}. 
It is worth noting that the \gls{DMRS} positions are fixed within a frame, however, the pilot symbols itself depend on the actual slot number within the frame.
The proposed neural receiver operates on a \emph{per-slot} basis, i.e., the received resource grid of 14 consecutive OFDM symbols which represent an entire transport block of 67584 information bits per layer for our configuration.

\begin{algorithm}[t]
	\SetAlgoLined
	\SetKwInOut{Input}{Input}
	\SetKwInOut{Output}{Output}
	\SetKwBlock{Repeat}{repeat}{}
	\SetKwFor{RepTimes}{For}{do}{end}
	\SetKwFor{RepTimesshort}{For}{}{}
	\DontPrintSemicolon
	\Input{
        Number of iterations $N_\text{it}$\\

        Number of layers $N_\text{T}$\\

        Received post-FFT signal $\Ym \in \mathbb{C}^{N_\text{F}, N_\text{S}, N_\text{RX}}$\\

        Pos. encoded pilot distance $\Pm_{n_\text{T}} \in \mathbb{R}^{N_\text{F}, N_\text{S}, 2}$\\ for each layer ${n_\text{T}}$\\

        [Optional] Noise power spectral density \\$\Nm_0 \in \mathbb{R}^{N_\text{F}, N_\text{S}, N_\text{RX}}$ in dB\\

    }
	\Output{Soft-estimate $\ellv \in \mathbb{R}^{N_\text{F}, N_\text{S}, N_\text{T}, m}$ for each (coded) bit for each layer}

    \RepTimesshort{$n_\text{T}=0, \dots, N_\text{T}-1$}{
        \# Initial LS estimation \& interpolation (see~(\ref{eq:ls}))\\
        $\hat{\Hm}_{n_\text{T}} \leftarrow \operatorname{LS\_estimate}(\Ym, n_\text{T})$\\

        \# Input Embedding (see Fig.~\ref{fig:cnn_init}) \\
            $\Sm_{n_\text{T}}^{(0)} \leftarrow \operatorname{CNN_{init}}(\Ym,\Pm_{n_\text{T}}, \Nm_0, \hat{\Hm}_{n_\text{T}})$\\
    }
	\# Run receiver iterations \\
	\RepTimesshort{$t=0, \dots,$ $N_\text{it}-1$}{
        \RepTimesshort{$n_\text{T}=0, \dots, N_\text{T}-1$}{
            \# Message passing \\
            $\mv'_{n_F, n_S, n_\text{T}} \leftarrow \text{MLP}_{\text{MP}}^{(t)} \LB \sv_{n_F, n_S, n_\text{T}}^{(t)} \RB \quad \forall n_\text{F}, n_\text{S}$\\

            \# Aggregate states (assuming $N_\text{T} \geq 2$)\\
            $\mv_{n_F, n_S, n_\text{T}}^{(t)} \leftarrow \frac{1}{N_\text{T}-1} \sum_{n_\text{T}' \neq n_\text{T}} \mv'_{n_\text{F}, n_\text{S}, n_\text{T}'} \, \forall n_\text{F}, n_\text{S}$\\

            \# Update states (see Fig.~\ref{fig:cnn_it}) \\
            $\Sm_{n_\text{T}}^{(t+1)} \leftarrow \operatorname{CNN}_\text{state}(\Mm_{n_\text{T}},\Pm_{n_\text{T}}, \Sm_{n_\text{T}}^{(t)}) $\\

        }

    }

    \# Read-out function \\
    \RepTimesshort{$n_\text{T}=0, \dots, N_\text{T}-1$}{
        $\ellv_{n_F, n_S, n_\text{T}} \leftarrow \operatorname{MLP_{readout}}(\Sm_{n_F, n_S, n_\text{T}}^{(N_\text{it})}) \quad \forall n_\text{F}, n_\text{S}$
    }
	\Return $\ellv$ 

	\caption{Neural \gls{MUMIMO} receiver}
	\label{alg:neural_rx}
\end{algorithm}

\subsection{Neural Receiver Architecture}
\vspace*{-0.1cm}
Alg.~\ref{alg:neural_rx} implements the neural receiver that is shown in Fig.~\ref{fig:nn}.
It processes an entire \emph{slot} and provides a one-shot estimation of all transmitted bits of all layers simultaneously.
The neural receiver takes as input:\footnote{The actual implementation is batched, i.e., multiple slots and layers are processed in parallel. However, for readability this is omitted in the following description.}
\begin{itemize}
    \item The complex-valued received \emph{post-FFT} signal $\Ym$ as a 3-dimensional tensor of shape $N_\text{F} \times N_\text{S} \times N_\text{RX}$.

    \item The real-valued \gls{PE} $\Pm_{n_\text{T}}$ that provides the normalized distance to the nearest pilot in time (\gls{OFDM} symbols) and frequency (subcarriers). This is provided as a 3-dimensional tensor of shape $N_\text{F} \times N_\text{S} \times 2$. As the pilot pattern is specific to each user, $\Pm_{n_\text{T}}$ is also user specific. The implementation is shown in Alg.~\ref{alg:pos_encoding}.

    \item Optionally, the real-valued noise power spectral density $\Nm_0$ as a tensor of shape $N_\text{F} \times N_\text{S} \times N_\text{RX}$ in \SI{}{\dB}. Logarithmic scale is used to facilitate operating over large ranges of \gls{SNR}. If not provided the neural receiver can also learn an implicit \gls{SNR} estimation.

    \item Optionally, a complex-valued channel estimate $\widehat{\Hm}_{n_\text{T}}$ of shape $N_\text{F} \times N_\text{S} \times N_\text{RX}$.
    A \gls{LS} estimate of the channel vector for layer $n_\text{T}$ can be obtained for every \gls{RE} carrying a pilot according to
    \begin{equation}
        \vspace*{-0.1cm}
        \widehat{\hv}_{n_F, n_S, n_\text{T}} = \frac{p_{n_F,n_S,n_\text{T}}^* \yv_{n_F,n_S}}{\abs{p_{n_F,n_S,n_\text{T}}}^2}
        \label{eq:ls}
    \end{equation}
    where $p_{n_F,n_S,n_\text{T}} \in \CC$ is the pilot signal transmitted by layer $n_\text{T}$ on the \gls{RE} $\left[ n_F, n_S \right]$.
    Linear interpolation may then be used to obtain estimates of the channel vectors for data-carrying \glspl{RE}.
    Feeding the receiver with an estimate of the channel coefficients was experimentally found to improve the performance, as it may serve as a bootstrap for channel estimation. This also helps in scenarios where multiple users share the same \gls{CDM} group and the initial channel estimation can simplify the initial layer separation. Further, the \gls{LS} estimator introduces only a slight increase in complexity.
\end{itemize}

The essential processing steps of Alg.~\ref{alg:neural_rx} consist of three stages: an initialization step, $N_\text{it}$ unrolled iterations, and a read-out step. During inference, the read-out step is only required in the last iteration.
We outline the three stages hereafter:

\begin{algorithm}[t]
	\SetAlgoLined
	\SetKwInOut{Input}{Input}
	\SetKwInOut{Output}{Output}
	\SetKwBlock{Repeat}{repeat}{}
	\SetKwFor{RepTimes}{For}{do}{end}
	\SetKwFor{RepTimesshort}{For}{}{}
	\DontPrintSemicolon
	\Input{Set of pilot positions $\mathcal{P}_{n_\text{T}}$ for layer $n_\text{T}$
	}
	\Output{Positional pilot encoding $\Pm_{n_\text{T}} \in \mathbb{R}^{N_\text{F}, N_\text{S}, 2}$}

	    \RepTimesshort{$n_\text{F}=1, \dots, N_\text{F}$}{
            \RepTimesshort{$n_\text{S}=1, \dots, N_\text{S}$}{

                \# Find closest pilot distance in time\\
                $\hat{p}_\text{S} \leftarrow \operatorname{min}_{\pv \in \mathcal{P}_{n_\text{T}}} \left( |p_{\text{S}} - n_\text{S}| \right)$\\
                \# Find closest pilot distance in frequency\\
                $\hat{p}_\text{F} \leftarrow \operatorname{min}_{\pv \in \mathcal{P}_{n_\text{T}}} \left( |p_{\text{F}} - n_\text{F}| \right)$\\

                \# Store distance in $\tilde \Pm$\\
                $\tilde \pv_{n_\text{F},n_\text{S}} \leftarrow \left[\hat{p}_\text{S}, \hat{p}_\text{F}\right]$
            }
        }
    \# Normalize to unit variance and zero mean
    $ \Pm_{n_\text{T}} \leftarrow \operatorname{norm} \left(\tilde \Pm \right)$\\
    \Return $\Pm_{n_\text{T}}$
	\caption{Positional pilot encoding}
	\label{alg:pos_encoding}
\end{algorithm}

\subsubsection*{Initialization step}

\begin{figure}[t]
    \center
    \resizebox{0.85\columnwidth}{!}{\begin{tikzpicture}

\tikzstyle{box} = [draw,rounded corners=.1cm,inner sep=5pt,minimum height=1.4em, text width=6.6em, align=center, thick]

\tikzstyle{boxsmall} = [draw,rounded corners=.1cm, minimum height=1.5em, text width=1.8em, align=center, thick]

\def\stringList{{Concatenate},{Separable Conv.},{Layer Norm.},{ReLU},{Separable Conv.},{Layer Norm.},{ReLU},{Separable Conv.},{Layer Norm.}}
\def\dist{0.75} 

\foreach \s [count=\xi from 1] in \stringList{
    \node[draw, box, rotate=90] (node\xi) at (\xi*\dist, 0) {\s};}

\def\off{-1.35}
\node[] (y) at (\off, 1.05) {$\Ym$};
\node[boxsmall] (y1) at (-0.1, 1.05) {$\mathbb{C}2\mathbb{R}$};
\node[] (pe) at (\off, 0.4) {$\Pm_{n_\text{T}}$};
\node[] (no) at (\off, -0.4) {$\Nm_0$};
\node[] (h) at (\off, -1.05) {$\hat{\Hm}_{n_{T}}$};
\node[boxsmall] (h1) at (-0.1, -1.05) {$\mathbb{C}2\mathbb{R}$};
\draw [->,thick] (y) -- (y1.west) {};
\draw [->,thick] (pe) -- ([yshift=0.4cm]node1.north) {};
\draw [->,thick] (no) -- ([yshift=-0.4cm]node1.north) {};
\draw [->,thick] (h) -- (h1.west) {};

\node[] (s) at (7.9,0) {$\Sm_{n_\text{T}^{(0)}}$};
\draw [->,thick] (node9.south) -- (s.west) {};

\end{tikzpicture}}
    \caption{Architecture of the CNN used to compute the initial state tensor for every user. All separable convolutional layers use kernels of size $3 \times 3$. The first two separable convolutional layers use 128 kernels, and the last one uses $d_S$ units.\label{fig:cnn_init}}

\end{figure}
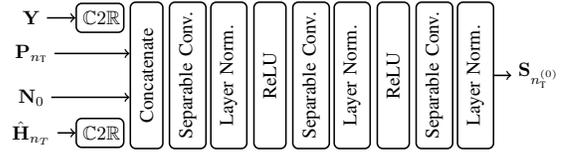

The initialization step consists of a small \gls{CNN}, whose architecture is shown in Figure~\ref{fig:cnn_init}.
The complex-valued inputs are converted to real-valued tensors by concatenating the real and imaginary components along the feature dimension.
Separable convolutions with \gls{relu} activations are used to reduce complexity.
The task of this \gls{CNN} is to compute an initial state tensor of shape $N_\text{F} \times N_\text{S} \times d_S$ for every layer $n_\text{T}$, denoted by $\Sm_{n_\text{T}}^{(0)}$ and the hyperparameter $d_S$ is the size of the feature dimension.
The state tensor of layer $n_\text{T}$ holds information about the channel and data for this layer and for every \gls{RE} $\left[ n_F, n_S \right]$.

\subsubsection*{Unrolled iterative algorithm}

The main stage of the proposed neural receiver consists of an unrolled iterative algorithm with $N_\text{it}$ iterations.
Each iteration updates the state tensor of every layer from $\Sm_{n_\text{T}}^{(t)}$ to $\Sm_{n_\text{T}}^{(t+1)}$, $0 \leq t \leq N_\text{it}-1$.
Each iteration consists of the following two steps.

\paragraph{Message passing}

The first step consists of message passing between all layers performed individually for every \gls{RE} $\left[ n_F, n_S \right]$.
Given an \gls{RE} $\left[ n_F, n_S \right]$, the state of each layer $n_\text{T}$ is computed using a \gls{MLP} according to
\begin{equation}
    \mv'_{n_F, n_S, n_\text{T}} = \text{MLP}^{(t)} \LB \sv_{n_F, n_S, n_\text{T}}^{(t)} \RB
    \label{eq:mlp_user}
\end{equation}
where $\sv_{n_F, n_S, n_\text{T}}^{(t)} \in \RR^{d_S}$ is the sub-vector of $\Sm^{(t)}_{n_\text{T}}$ corresponding to the \gls{RE} $\left[ n_F, n_S \right]$ and $\mv' \in \mathbb{R}^{d_\text{M}}$.
One may consider $\sv_{n_F, n_S, n_\text{T}}^{(t)}$ as carrying the data and channel information for the \gls{RE} $\left[ n_F, n_S \right]$ of layer $n_\text{T}$, where the hyperparameter $d_M$ is the size of the messages. 
The \gls{MLP} has a single hidden layer with \gls{relu} activations.
Note that the same \gls{MLP} is used for all pairs of layers and for all \glspl{RE}.

The messages intended for layer $n_\text{T}$ are then aggregated as
\begin{equation}
    \mv_{n_F, n_S, n_\text{T}} =
        \begin{cases}
        \frac{1}{N_\text{T}-1} \sum_{n_\text{T}' \neq n_\text{T}} \mv'_{n_F, n_S, n_\text{T}'} & \text{ if } N_\text{T} \geq 2\\
        \zerov & \text{ if } N_\text{T} = 1
    \end{cases}
\end{equation}
to form the message tensor $\Mm_{n_\text{T}}^{(t)}$ of shape $N_\text{F} \times N_\text{S} \times d_M$.

\paragraph{State update}

\begin{figure}[t]
    \center
    \resizebox{0.85\columnwidth}{!}{
        \begin{tikzpicture}

\tikzstyle{box} = [draw,rounded corners=.1cm,inner sep=5pt,minimum height=1.4em, text width=6.6em, align=center, thick]

\tikzstyle{boxsmall} = [draw,rounded corners=.1cm, minimum height=1.5em, text width=1.8em, align=center, thick]

\def\stringList{{Concatenate},{Separable Conv.},{Layer Norm.},{ReLU},{Separable Conv.},{Layer Norm.},{ReLU},{Separable Conv.}}
\def\dist{0.75} 

\foreach \s [count=\xi from 1] in \stringList{
    \node[draw, box, rotate=90] (node\xi) at (\xi*\dist, 0) {\s};}

\node[draw, circle, thick, minimum size=0.1cm] (resnet) at (9*\dist+0.25, 0.) {$+$};
\node[draw, box, rotate=90] (node9) at (10*\dist+0.5, 0) {Layer Norm.};

\draw [->,thick] (node8.south) -- (resnet.west) {};
\draw [->,thick] (resnet.east) -- (node9.north) {};

\def\off{-0.7}
\node[] (pe) at (\off, 0.8) {$\Pm_{n_\text{T}}$};
\node[] (no) at (\off, 0.) {$\Mm_{n_\text{T}}^{(t)}$};
\node[] (s) at (\off, -0.8) {$\Sm_{n_\text{T}}^{(t)}$};

\draw [->,thick] (pe) -- ([yshift=0.8cm]node1.north) {};
\draw [->,thick] (no) -- ([yshift=0.cm]node1.north) {};
\draw [->,thick] (s) -- ([yshift=-0.8cm]node1.north) {};

\node[] (s) at (9.3,0) {$\Sm_{n_\text{T}}^{(t+1)}$};
\draw [->,thick] (node9.south) -- (s.west) {};

\node[] (A) at (-0.05, -0.68) {};
\node[] (B) at (5, -1.6) {};
\node[] (B1) at (4.75, -1.6) {};
\draw[thick, -] (A) |- (B);
\draw[thick, ->] (B1) -| (resnet.south);

\end{tikzpicture}
        }
    \caption{Architecture of the CNN used to update the state of every layer. All separable convolutional layers use kernels of size $3 \times 3$. The first two separable convolutional layers use 256 kernels, and the last one uses $d_S$ kernels. 
    \label{fig:cnn_it}}
\end{figure}
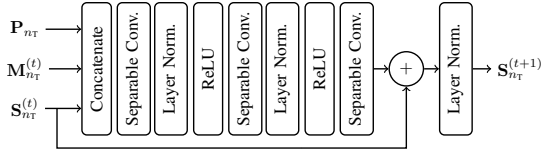

The second step of every iteration consists in updating the state tensor of every layer $n_\text{T}$ from the message tensor $\Mm_{n_\text{T}}^{(t)}$, the current state tensor $\Sm_{n_\text{T}}^{(t)}$, and the \gls{PE} $\Pm_{n_\text{T}}$.
This is implemented independently for all layers, using a \gls{CNN} whose architecture is shown in Fig.~\ref{fig:cnn_it}.
A skip connection is used to avoid vanishing gradients, and separable convolutions with \gls{relu} activations to reduce complexity.
The same \gls{CNN} is used for all layers.
Using a \gls{CNN} enables exploiting the time-frequency correlation of the channel coefficients, while taking into account the interference from the other users through the message tensor.
Weight sharing across iterations was found to significantly degrade the performance and hence cannot be used.

\subsubsection*{Read-out}

The last stage consists of computing \glspl{LLR} $\ellv_{n_F, n_S, n_\text{T}}^{(t)} \in \mathbb{R}^m$ for the $m$ bits $\bv_{n_F, n_S, n_\text{T}}$ transmitted by the layer over this \gls{RE} for every layer $n_\text{T}$ and every \gls{RE} $\left[ n_F, n_S \right]$.
This is achieved using a read-out network implemented by an \gls{MLP}, that individually processes the state vectors $\sv^{(t)}_{n_F, n_S, n_\text{T}}$ of every \gls{RE} and layer, such that
$$\ellv_{n_F, n_S, n_\text{T}}^{(t)} = \operatorname{MLP_{readout}}(\Sm_{n_F, n_S, n_\text{T}}^{(t)}) \quad \forall n_\text{F}, n_\text{S}, n_\text{T}.$$
Note that the same \glspl{MLP} are used for all layers and for all \glspl{RE}.
As a result, the entire network consists of approximately 730k trainable weights.

\subsection{Training \& Generalization}

\begin{algorithm}[tb]
	\SetAlgoLined
	\SetKwInOut{Input}{Input}
	\SetKwInOut{Output}{Output}
	\SetKwBlock{Repeat}{repeat}{}
	\SetKwFor{RepTimes}{For}{do}{end}
	\SetKwFor{RepTimesshort}{For}{}{}
	\DontPrintSemicolon
	\Input{
        Batch size $B$, learning rate $l$\\
        Number training iterations $N_\text{train,it}$\\
        SNR range $[\sigma^2_{\text{min}}, \sigma^2_{\text{max}}]$ \\ 
        Max. number of layer $N_{\text{T,max}}$\\
        Initial weights $\thetav^{(0)}$\\
    }
	\Output{Optimized weights $\thetav^*$}

    \# Apply $N_\text{train,it}$ SGD training iterations\\
    \RepTimesshort{$t=0, \dots, N_\text{train,it}-1$}{
        \# Sample random number of layers\\
        $N_\text{T} \leftarrow$  triangleSampler(1, $N_{\text{T,max}}$)\\

        $\mathcal{L} \leftarrow 0$ \qquad\# Average loss over batch\\

        \RepTimesshort{$0, \dots, B$}{
            \# Sample random SNR\\
            $\sigma^2 \leftarrow$  SNRSampler($\sigma^2_{\text{min}}, \sigma^2_{\text{max}}$)\\

            \# Draw new random channel realization\\
            $\Hm \leftarrow $ getChannel($N_\text{T}$)\\

            \# Sample random bits $\Bm$ and simulate transmission \\
            $\Ym, \Bm \leftarrow$  simTransmission($\Hm$, $\sigma^2$, $N_\text{T}$)\\

            \# Apply neural receiver\\
            $\ellv \leftarrow$ runNeuralRX($\Ym$, $\thetav^{(t)}$, $N_\text{T}$)\\

            \# Calculate and accumulate BCE loss from (\ref{eq:loss})\\
            $\mathcal{L} \leftarrow \mathcal{L}$ + calcLoss($\ellv$, $\Bm$)\\

        }
        \# Apply SGD and update weights\\
        $\thetav^{(t+1)} \leftarrow$ SGD($\mathcal{L}/B$, $\thetav^{(t)}$, $l$)\\
    }

	\Return $\thetav^* \leftarrow \thetav^{N_\text{train,it}}$
	\caption{Training Loop}
	\label{alg:training}

\end{algorithm}

The proposed neural receiver architecture is fully differentiable and, hence, \gls{SGD}-based training is straightforward. We use the \gls{ADAM} optimizer with learning rate $l=10^{-3}$, and the \gls{BCE} loss function. Thus, the neural receiver can be seen as attempting to solve $N_\text{F} \times N_\text{S} \times N_\text{T} \times m$ binary classification problems\footnote{For readability, we ignore the fact that some \glspl{RE} are allocated to pilots in the following description, however, in the actual implementation these positions are excluded from the calculation of the loss.} in parallel from the received signal $\Ym$. The detailed training algorithm can be found in Alg.~\ref{alg:training}.
We can estimate the \gls{BCE} by Monte-Carlo integration
\begin{multline}
    \label{eq:loss}
    \begin{split}
    \mathcal{L} & \approx  -\frac{1}{B N_\text{F} N_\text{S} N_\text{T} m }\sum_{b = 0}^{B-1} \sum_{n_F = 1}^{N_\text{F}} \sum_{n_S = 1}^{N_\text{S}} \sum_{n_\text{T} = 0}^{N_\text{T}-1} \sum_{i = 0}^{m-1} \\
    & \Big[ b_{n_F,n_S,n_\text{T},i}^{[b]} \log{\sigma\LB  \ell_{n_F,n_S,n_\text{T},i} \LB \Ym^{[b]} \RB\RB} \\
    & + \LB 1 - b_{n_F,n_S,n_\text{T},i}^{[b]} \RB \log{\sigma\LB-\ell_{n_F,n_S,n_\text{T},i} \LB\Ym^{[b]}\RB\RB} \Big]
    \end{split}
\end{multline}

where $B$ is the batch size, the superscript $[b]$ is used to refer to the $b^{\text{th}}$ batch example, and $\sigma(\cdot)$ is the logistic sigmoid function.
$\ell_{n_F,n_S,n_\text{T},i}(\Ym)$ denotes the \gls{LLR} computed by the detector from the received signal $\Ym$ for the $i^{\text{th}}$ bit of layer $n_\text{T}$ transmitted over the \gls{RE} $\left[ n_F, n_S \right]$. As \glspl{LLR} are binary logits, $\sigma \LB \ell_{n_F,n_S,n_\text{T},i}(\Ym) \RB$ gives the corresponding probability for the bit to be equal to one given $\Ym$.
For training, we average the loss over all iterations to ensure that inference of the receiver can be done for a flexible number of iterations (see \emph{multi-loss} in \cite{nachmani2016learning}).
Moreover, training on the \gls{BCE} was shown to be equivalent to minimizing the \gls{KL} divergence between the posterior distribution on the bits approximated by the neural receiver and the true one that would be given by an optimal receiver (cf.~\cite{stark2019joint}).
Further, the \glspl{LLR} after channel decoding can be used for training if the channel decoder is differentiable (e.g., for \gls{BP} decoding \cite{nachmani2016learning}). However, we empirically did not observe any gains by doing so.

To account for a flexible number of layers, we randomly sample a different number of active users for every batch, i.e., per \gls{SGD} iteration.
Intuitively, the task of \gls{MUMIMO} detection becomes more complicated if a larger number of users is active and, therefore, we use a triangular distribution for the number of active layers \cite{9298921} such that scenarios with many active layers occur more frequently.

To avoid overfitting to a particular channel condition, the neural receiver was trained on the 3GPP \gls{UMi} model built-in Sionna \cite{hoydis2022sionna}, with a random drop of users for every batch example. This ensures that the \gls{NN} does not overfit to specific power delay profiles and angles of arrival and departure.
Moreover, the user speeds are independently and uniformly sampled from the range $[0,34]$ \SI{}{\m\per\s} at training.
This procedure ensures that the resulting neural receiver generalizes well to unseen channel realizations. For training with real world data, one can also embed actual measurements in the synthetic training dataset to ensure that the receiver still generalizes as the acquisition of samples from many different channel measurements turns out to be cumbersome and expensive.

\section{Results Evaluation}

For our experiments, we use the Sionna~\cite{hoydis2022sionna} link-level simulator to train and evaluate the neural receiver. 
For all experiments, we focus on the \gls{5GNR} \gls{PUSCH} scenario and set $N_\text{S} = 14$ \gls{OFDM} symbols per slot (see Sec.~\ref{sec:5gnr} for details). For the training, the number of \glspl{PRB} is set to 4 (i.e., $N_\text{F} = 48$ sub-carriers) and 217 \glspl{PRB} (i,e., $N_\text{F} = 2604$ sub-carriers) for the evaluation, respectively. This underlines that the neural receiver shall support a variable number of subcarriers. The carrier frequency is set to \SI{2.14}{GHz} and the subcarrier spacing is \SI{30}{kHz}.
We use the 3GPP \gls{UMi} model \cite{38901} with randomized user positions and user speeds between \SI{0}{\m\per\s} and \SI{34}{\m\per\s}.
Further, the \gls{MCS} index is set to 14, leading to Gray labelled 16-\gls{QAM}, and  \gls{LDPC} code of rate $0.54$ throughout all experiments.
Note that the code rate can be changed without the need for retraining, however, the \gls{QAM} constellation is \emph{baked} into the weights of the model. The same holds for the sub-carrier spacing, so that changing from \SI{30}{kHz} to \SI{60}{kHz} requires retraining of the current network.

\subsection{Varying number of layers}

\begin{figure*}
    \centering
    \pgfplotsset{compat=1.5}
\tikzset{font={\fontsize{8pt}{8}\selectfont}}

\def\figheight{4.1}
\def\figwidth{0.245\textwidth}
\def\vspacedist{-0.2}

\ref{jointLegend}

\vspace*{-0.1cm}
\begin{subfigure}[b]{\figwidth}
	\begin{tikzpicture}
		\begin{axis}[
            legend to name=jointLegend,  
			legend columns=-1,
			legend style={line width=1pt,nodes={scale=0.93, transform shape}},
			xmode=normal,
			ymode=log,
			xlabel=$E_b/N_0~(\mathrm{dB})$,
			ylabel=$\mathrm{TBLER}$,
			y label style={at={(axis description cs:-0.3,.5)},anchor=south},
			x label style={at={(axis description cs:0.5,-0.1)},anchor=north},
			xmin = -8,
			xmax = 4,
			ymax = 1e-0,
			ymin = 2.e-3,
			mark size=1.5pt,
			grid=both,
			minor grid style={gray!25},
			major grid style={gray!25},
			width=\columnwidth,
			height=\figheight cm,
			legend cell align={left},
			line width=1pt]

		\addplot+ [anthrazit, mark options={fill=anthrazit}]
		table[x=snr, y=bler, col sep=comma]{figs/data/cdm4_1_perfcsi.csv};
		\addlegendentry{Baseline - Perfect-CSI + K-Best Det.}

		\addplot+ [lila, mark repeat=2,mark phase=2, mark options={fill=lila}]
		table[x=snr, y=bler, col sep=comma]{figs/data/cdm4_1_lmmse+kbest.csv};
		\addlegendentry{Baseline - LMMSE Chest. + K-Best Det.}
		\addplot+ [orange, mark repeat=2, mark phase=2,mark options={fill=orange}]
		table[x=snr, y=bler, col sep=comma]{figs/data/cdm4_1_lslin+lmmse.csv};
		\addlegendentry{Baseline - LS Chest. + LMMSE Det.}
		\addplot+ [apfelgruen, mark options={fill=apfelgruen}]
		table[x=snr, y=bler, col sep=comma]{figs/data/cdm4_1_nn.csv};
		\addlegendentry{Neural Receiver}
	\end{axis}
\end{tikzpicture}
\vspace*{\vspacedist cm}
\label{fig:ber_nt1}
\caption{$N_T=1$}
\end{subfigure}
\begin{subfigure}[b]{\figwidth}
	\begin{tikzpicture}
	\begin{axis}[
		xmode=normal,
		ymode=log,
		xlabel=$E_b/N_0~(\mathrm{dB})$,
		ylabel=$\mathrm{TBLER}$,
		y label style={at={(axis description cs:-0.3,.5)},anchor=south},
		x label style={at={(axis description cs:0.5,-0.1)},anchor=north},
        xmin = -8,
        xmax = 4,
        ymax = 1e-0,
        ymin = 2.e-3,
		mark size=1.5pt,
		grid=both,
		minor grid style={gray!25},
		major grid style={gray!25},
		width=\columnwidth,
		height=\figheight cm,
		line width=1pt,
		]

		\addplot+ [anthrazit, mark options={fill=anthrazit}]
		table[x=snr, y=bler, col sep=comma]{figs/data/cdm4_2_perfcsi.csv};

		\addplot+ [lila, mark options={fill=lila}, mark indices={1,3,5,7,9,11,14,15}]
		table[x=snr, y=bler, col sep=comma]{figs/data/cdm4_2_lmmse+kbest.csv};

		\addplot+ [orange, mark repeat=2, mark phase=2,mark options={fill=orange}]
		table[x=snr, y=bler, col sep=comma]{figs/data/cdm4_2_lslin+lmmse.csv};

		\addplot+ [apfelgruen, mark options={fill=apfelgruen}]
		table[x=snr, y=bler, col sep=comma]{figs/data/cdm4_2_nn.csv};

	\end{axis}
	\end{tikzpicture}
	\vspace*{\vspacedist cm}
	\label{fig:ber_nt2}
	\subcaption{$N_T=2$}
\end{subfigure}
\begin{subfigure}[b]{\figwidth}
	\begin{tikzpicture}
	\begin{axis}[
		xmode=normal,
		ymode=log,
		xlabel=$E_b/N_0~(\mathrm{dB})$,
		ylabel=$\mathrm{TBLER}$,
		y label style={at={(axis description cs:-0.3,.5)},anchor=south},
		x label style={at={(axis description cs:0.5,-0.1)},anchor=north},
        xmin = -8,
        xmax = 4,
        ymax = 1e-0,
        ymin = 2.e-3,
		mark size=1.5pt,
		grid=both,
		minor grid style={gray!25},
		major grid style={gray!25},
		width=\columnwidth,
		height=\figheight cm,
		line width=1pt]

		\addplot+ [anthrazit, mark options={fill=anthrazit}]
		table[x=snr, y=bler, col sep=comma]{figs/data/cdm4_3_perfcsi.csv};

		\addplot+ [lila, mark options={fill=lila}, mark indices={1,3,5,7,9,11,14,15}]
		table[x=snr, y=bler, col sep=comma]{figs/data/cdm4_3_lmmse+kbest.csv};

		\addplot+ [orange, mark repeat=2, mark phase=2, mark options={fill=orange}]
		table[x=snr, y=bler, col sep=comma]{figs/data/cdm4_3_lslin+lmmse.csv};

		\addplot+ [apfelgruen, mark options={fill=apfelgruen}]
		table[x=snr, y=bler, col sep=comma]{figs/data/cdm4_3_nn.csv};

	\end{axis}
	\end{tikzpicture}
	\label{fig:ber_nt3}
	\vspace*{\vspacedist cm}
	\caption{$N_T=3$}
\end{subfigure}
\begin{subfigure}[b]{\figwidth}
	\begin{tikzpicture}
	\begin{axis}[
		xmode=normal,
		ymode=log,
		xlabel=$E_b/N_0~(\mathrm{dB})$,
		ylabel=$\mathrm{TBLER}$,
		y label style={at={(axis description cs:-0.3,.5)},anchor=south},
		x label style={at={(axis description cs:0.5,-0.1)},anchor=north},
        xmin = -8,
        xmax = 4,
        ymax = 1e-0,
        ymin = 2.e-3,
		mark size=1.5pt,
		grid=both,
		minor grid style={gray!25},
		major grid style={gray!25},
		width=\columnwidth,
		height=\figheight cm,
		line width=1pt]

		\addplot+ [anthrazit, mark options={fill=anthrazit}]
		table[x=snr, y=bler, col sep=comma]{figs/data/cdm4_4_perfcsi.csv};

		\addplot+ [lila, mark options={fill=lila}, mark indices={0,2,4,6,8,10,12,15,17}]
		table[x=snr, y=bler, col sep=comma]{figs/data/cdm4_4_lmmse+kbest.csv};

		\addplot+ [orange, mark repeat=2, mark phase=2,mark options={fill=orange}]
		table[x=snr, y=bler, col sep=comma]{figs/data/cdm4_4_lslin+lmmse.csv};

		\addplot+ [apfelgruen, mark options={fill=apfelgruen}]
		table[x=snr, y=bler, col sep=comma]{figs/data/cdm4_4_nn.csv};

	\end{axis}
	\end{tikzpicture}
	\label{fig:ber_nt4}
	\vspace*{\vspacedist cm}
	\caption{$N_T=4$}
\end{subfigure}
    \vspace*{-0.4cm}
    \caption{Varying number of active layers $N_\text{T}$ for a 3GPP UMi channel model and $N_\text{RX}=16$ receive antennas. The receiver was trained for $N_{\text{T,max}}=4$. We randomize for every sample the user positions and user mobility from 0 to \SI{34}{\m\per\s}.}
    \label{fig:results_user}
    \vspace*{-0.6cm}
\end{figure*}
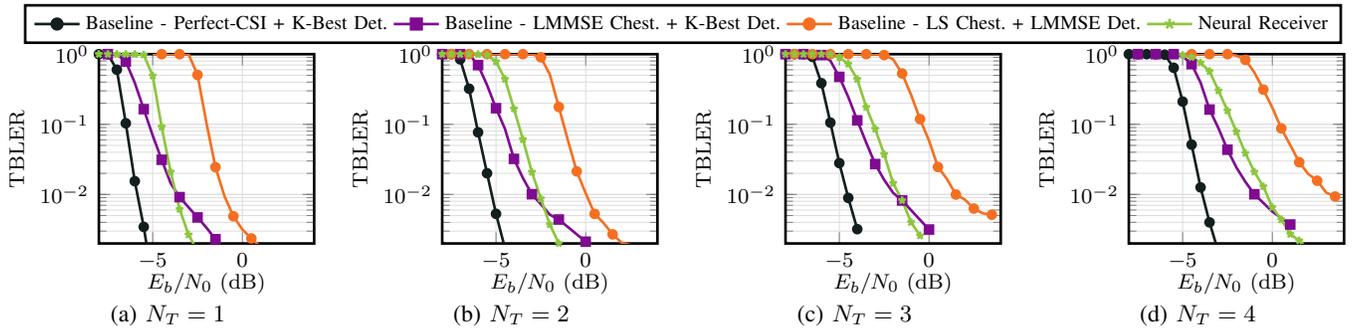

One of the key challenges of neural networks is the possibility to re-parametrize the network without the need for retraining.
The number of receive antennas is set to $N_\text{RX} = 16$, and the neural receiver is trained with a randomly sampled number of active layers, $N_\text{T}$, ranging from 1 to 4. Note that in our configuration, there are only two \gls{CDM} groups and two users must share the same \gls{CDM} group, i.e., their pilots are transmitted on the same resource elements.

The results in Fig.~\ref{fig:results_user} show the \gls{TBLER} evaluation for the \emph{same} neural receiver for a different number of active users without any additional retraining.
We compare the neural receiver to a conventional baseline using \gls{LS} channel estimation at pilot locations and linear interpolation across the data symbols, and \gls{LMMSE} equalization.
Further, we show a baseline using \gls{LMMSE} channel estimation with K-best detection which has a high computational complexity, but can be considered as an approximation of maximum likelihood detection.
As one can see, the proposed architecture outperforms the \gls{LS} baseline in all scenarios and operates at a performance close to the baseline using \gls{LMMSE} channel estimation with K-best detection, but at a significantly lower computational complexity. Further, the error floor is significantly lower with the neural receiver, suggesting that it implicitly performs a more accurate channel estimation than the two other schemes.

\subsection{Hardware-in-the-loop \& 3GPP conformance test}

As a final experiment, we validate our implementation with 5G NR \gls{MUMIMO} RF signals in a hardware-in-the-loop setup based on the R\&S®SMW200A vector signal generator, the R\&S®MSR4 multi-purpose (satellite) receiver, and a signal analysis software framework to pre-process the received signals (synchronization, FFT calculation). With this setup, we emulate two users ($N_\text{T}=2$), each equipped with two transmit antennas, and use the four inputs of the receiver with $N_\text{RX}=4$ antennas. We use 217 \glspl{PRB} for the hardware setup. To account for the different channels of both users at different positions, we emulate two different \gls{TDL} models from the  \gls{3GPP} 38.901 conformance test cases in the signal generator:

\begin{itemize}
     \item TDL-B100/400: TDL-B with 100ns delay spread and 400Hz Doppler shift \cite{38901}
     \item TDL-C300/100: TDL-C with 300ns delay spread and 100Hz Doppler shift \cite{38901}
\end{itemize}
The receiver \emph{sees} the superposition of the two channel outputs with an additional \gls{AWGN} component.
We want to emphasize that the receiver is \emph{not} trained for this specific channel model and, thus, can operate on any random realization of the entire \emph{ensemble} of possible \gls{UMi} channels.

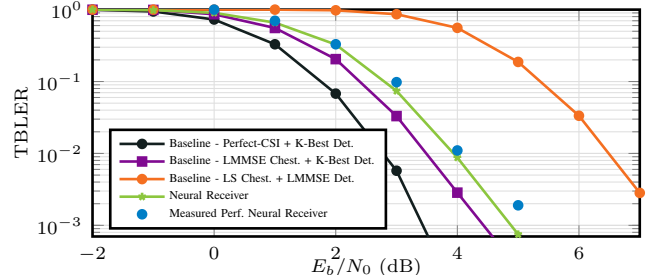
\begin{figure}
    \center
    \vspace*{-0.3cm}
    \begin{tikzpicture}
	\pgfplotsset{compat=1.5}
	\tikzset{font={\fontsize{7pt}{7}\selectfont}}
	\begin{axis}[
		xmode=normal,
		ymode=log,
		xlabel=$E_b/N_0~(\mathrm{dB})$,
		ylabel=$\mathrm{TBLER}$,
		y label style={at={(axis description cs:-0.1,.5)},anchor=south},
		x label style={at={(axis description cs:0.5,-0.05)},anchor=north},
		xmin = -2,
		xmax = 7,
		ymax = 1e-0,
		ymin = 7e-4,
		mark size=1.5pt,
		legend style={nodes={scale=0.65, transform shape}},
		legend pos = south west,
		grid=both,
		minor grid style={gray!25},
		major grid style={gray!25},
		width=\columnwidth,
		height=4.6cm,
		legend cell align={left},
		line width=1pt]

		\addplot+ [anthrazit, mark options={fill=anthrazit}]
		table[x=snr, y=perfcsi, col sep=comma]{figs/mwc_results.txt};
		\addlegendentry{Baseline - Perfect-CSI + K-Best Det.}

		\addplot+ [lila, mark options={fill=lila}]
		table[x=snr, y=ml, col sep=comma]{figs/mwc_results.txt};
		\addlegendentry{Baseline - LMMSE Chest. + K-Best Det.}

		\addplot+ [orange, mark options={fill=orange}]
		table[x=snr, y=baseline, col sep=comma]{figs/mwc_results.txt};
		\addlegendentry{Baseline - LS Chest. + LMMSE Det.}

		\addplot+ [apfelgruen, mark options={fill=apfelgruen}]
		table[x=snr, y=mwc, col sep=comma]{figs/mwc_results.txt};
		\addlegendentry{Neural Receiver}

		\addplot[only marks, mark size=1.5, mittelblau, mark options={color=mittelblau, fill=mittelblau}]
		coordinates{
			(0,1)
			(1, 0.7)
			(2, 0.33)
			(3, 0.098)
			(4, 0.011)
			(5, 0.0019)};
		\addlegendentry{Measured Perf. Neural Receiver}

	\end{axis}
\end{tikzpicture}
    \vspace*{-0.3cm}
    \caption{Hardware-in-the-loop results for $N_\text{T}=2$ user \gls{MIMO} with $N_\text{RX}=4$ receiver antennas and TDL-B100/400 \& TDL-C300/100 channels. No retraining for the TDL channel is done.}
    \label{fig:results_mwc}
\end{figure}

Fig.~\ref{fig:results_mwc} shows the \gls{BLER} at transport block level including all physical layer effects of an entire slot.
As can be seen, the neural receiver operates close to the baseline consisting of \gls{LMMSE} channel estimation with K-best detection but at a significantly lower computational complexity. Further, the neural receiver outperforms the more practical receiver using \gls{LS} channel estimation and \gls{LMMSE}-based detection significantly. While the solid green curve is a simulated result, the blue dots indicate the measured \gls{TBLER} for the actual experiment. The small performance mismatch can be explained by the aggregation of all hardware effects (quantization noise, carrier frequency offset, ...) which are not considered in the simulations. Further, slightly different numerical implementations of the channel models are possible.

\section{Conclusion}

We have presented a neural \gls{MUMIMO} \gls{OFDM} receiver and demonstrated its \gls{5GNR} \gls{PUSCH} compatibility. The proposed receiver outperforms our classical baselines and is flexible with respect to the number of layers and \glspl{PRB}.
Beyond pure block error-rate performance gains, our plan is that these receivers can be finetuned for a specific environment considering the site-specific properties such as the expected user speed (e.g., highway vs. indoor scenarios) or the expected maximum delay spread for the area of interest. As such, we envision base stations that are continuously retrained---during low-load phases---and, thereby, improve their performance even after deployment in the field by simple weight updates.

Going further, neural receivers can be seen as an enabler for a plethora of new physical layer concepts such as \gls{AI}/\gls{ML}-based waveforms \cite{o2017introduction,aoudia2021end} or even semantic communications.
We leave it open for future work to optimize the architecture towards real-time inference. The flexibility w.r.t. the modulation schemes is an interesting and important challenge.
A possible approach for such a flexibility could be in input embeddings that account for the actual modulation scheme.

\bibliographystyle{IEEEtran}
\bibliography{IEEEabrv,bibliography}

\end{document}